\documentclass[aps,prl,preprint,showpacs,longbibliography,groupedaddress]{revtex4-1}

\usepackage{color,amsmath,amssymb,graphicx}

\begin{document}


\title{Synthesis of passive lossless metasurfaces using auxiliary fields for reflectionless beam splitting and perfect reflection}


\author{Ariel Epstein}
\email[]{ariel.epstein@utoronto.ca}
\author{George V. Eleftheriades}
\email[]{gelefth@waves.utoronto.ca}
\affiliation{The Edward S. Rogers Sr. Department of Electrical and
Computer Engineering, University of Toronto, Toronto, ON M5S 2E4 Canada}


\date{June 14, 2016}

\begin{abstract}
We introduce a paradigm for accurate design of metasurfaces for intricate beam manipulation, implementing functionalities previously considered impossible to achieve with passive lossless elements. The key concept involves self-generation of auxiliary evanescent fields which facilitate the required local power conservation, without interfering with the device performance in the far field. We demonstrate our scheme by presenting exact reactive solutions to the challenging problems of reflectionless beam splitting and perfect reflection, verified via full wave simulations.    
\end{abstract}


\maketitle

Thin planar arrangements of polarizable subwavelength particles (metasurfaces) have been attracting significant attention lately due to their demonstrated ability to efficiently implement a variety of electromagnetic functionalities \cite{Holloway2012, Tretyakov2015, Achouri2015_1, Estakhri2016, Epstein2016_2}. Many innovative low-profile devices manipulating the phase, magnitude, and polarization of aperture fields have been reported in recent years \cite{Yu2011, Selvanayagam2013, Monticone2013, Pfeiffer2013,Maci2011,Aieta2015,Pfeiffer2014_1,Pfeiffer2014_2,Asadchy2015,Asadchy2015_1,Epstein2016}, pointing out the immense potential of these surfaces in optics and microwave physics and engineering.

While the first demonstrations used a single layer of electrically polarizable particles (meta-atoms) \cite{Yu2011}, it was soon found that these suffer from a limited transmission efficiency. To achieve a unity transmission magnitude with arbitrary transmission phase, both electrically and magnetically polarizable meta-atoms must be used \cite{Monticone2013, Pfeiffer2013, Selvanayagam2013_2}. When illuminated by an incident field, such metasurfaces induce orthogonal electric and (equivalent) magnetic currents on the surface, forming Huygens' sources, capable of unidirectional radiation (i.e., full transmission) \cite{Pfeiffer2013}. This phenomenon can be fully accounted for by the generalized sheet transition conditions (GSTCs) \cite{Holloway2012, Tretyakov2015, Achouri2015_1, Estakhri2016, Epstein2016_2}, 
formulating the equivalent boundary conditions posed by the metasurface.
These so-called Huygens' metasurfaces (HMSs) inspired a large number of reports, utilizing unit cells with collocated electric and magnetic polarizability to achieve full transmission with a given phase profile (\textit{cf.} \cite{Epstein2016_2}).

Nevertheless, it was recently recognized that this common phase-stipulation design scheme is not accurate \cite{Estakhri2016, Epstein2016_2}. The basic problem lies in the fact that Maxwell's equations demand that any modification of the phase profile on an aperture will be accompanied by a suitable change in the field amplitudes, due to the inevitable variation of the wave impedance \cite{Epstein2014}. Thus, stipulating transmitted fields with uniform (unity) amplitude and arbitrary phase is not valid.

This recognition has dramatic implications. For instance, when the GSTCs for passive lossless HMSs are rigorously examined, it turns out that plane-wave refraction cannot in general be realized with zero reflection, as the metasurface \emph{symmetric} structure can only be matched to a single wave impedance value \cite{Selvanayagam2013, Epstein2014_2}. Indeed, the formulation in \cite{Epstein2014} showed that sufficient conditions to ensure passive and lossless design in this case include local power conservation and local wave-impedance equalization, requiring at least one additional diffraction mode, e.g. specular reflection. Although this reflection is minor for small-to-moderate angle refraction, it becomes significant when wide-angle refraction is considered.

These observations raised serious questions regarding previously-reported achievements. Primarily, could passive lossless metasurfaces achieve truly-reflectionless engineered refraction? In a recent paper, Wong \textit{et al.} \cite{Wong2016} provided a positive answer to that question. Relying on generalized scattering matrix theory, the authors showed that an \emph{asymmetric} stacking of three reactance sheets could be designed to simultaneously match the wave impedance of both the incident and refracted plane waves, leading to zero reflections even in the case of wide-angle refraction. This approach was generalized in \cite{Epstein2016_3}, showing that this structure corresponds to an omega-type bianisotropic metasurface (O-BMS). In O-BMSs, meta-atoms exhibit electric and magnetic polarizabilities as in HMSs, but also feature magnetoelectric coupling \cite{Tretyakov2015, Radi2013}. This additional degree of freedom allows passive and lossless implementation of metasurfaces which are matched to different wave impedances on their top and bottom facets; similarly, O-BMSs can be used for achieving asymmetric reflection properties \cite{Radi2014_1}. 

The wave impedance perspective, thus, can be used to find a passive and lossless solution for transforming a given incident plane wave to a desirable transmitted plane wave. But what happens when more than one mode exists below or above the metasurface? When the incident plane wave is to be splitted into two transmitted beams, or reflected in a different direction, the interference between the modes does not allow a proper definition of two distinctive network ports for the top and bottom metasurface facets; this prevents utilization of microwave network theory as in \cite{Wong2016}.
In fact, these fundamental diffraction engineering problems have very recently been examined by several authors, which either concluded that inclusion of lossy or active meta-atoms is required for the implementation, or suggested an approximate approach which does not form a valid solution to Maxwell's equations \cite{Asadchy2016, Estakhri2016}. To the best of our knowledge, whether or not an accurate design of passive lossless metasurfaces for such applications is possible remains an open question.

In this Letter, we propose a different paradigm to tackle these problems. The solution relies on the theorem proved in \cite{Epstein2016_3}, indicating that \emph{any} field transformation that \emph{locally}-conserves the real power can be implemented by passive and lossless O-BMSs. Once this is established, designing a reflectionless refracting metasurface becomes trivial, as both the incident and transmitted plane waves feature a uniform power profile on the metasurface; local power conservation is thus achieved by proper stipulation of the field amplitudes \cite{Wong2016, Epstein2016_3} (this result was independently derived in \cite{Asadchy2016}). On the other hand, trying to apply this theorem naively for the cases of beam splitting or engineered reflection will fail, as the simple stipulation of the incident and diffracted fields forms non-trivial spatially-varying power profiles on the metasurface, which do not match, in general. 

Hence, the key concept underlying our solution is the use of auxiliary fields to achieve local power conservation. This approach dictates a judicious stipulation of evanescent field components, additional to the prescribed excitation and desirable scattered fields, such that the interference between all modes guarantees the continuity of the real power at each point on the metasurface plane, without affecting the overall functionality of the metasurface device in the far field. The electromagnetic fields analytically stipulated by this scheme form an \emph{exact} solution to Maxwell's equations, 
supported by a suitable passive and lossless O-BMS design.

To demonstrate the versatility of our concept we present solutions to two basic diffraction engineering problems, namely, wide-angle beam splitting 
and engineered plane-wave reflection, featuring qualitatively-different power profiles. The designed metasurfaces are implemented in a commercially-available finite-element solver (ANSYS HFSS) following the scheme introduced in \cite{Wong2016, Epstein2016_3}, showing excellent agreement between the analytical predictions and full-wave simulations. It should be stressed that the physical rationale explored herein is general, and can be applied to a wide range of metasurface problems. The applications addressed in the paper were chosen to emphasize the conceptual leap required to implement challenging functionalities via passive lossless metasurfaces, deliberately introducing auxiliary fields to facilitate the solution.

\begin{figure*}
 \includegraphics[width=16.5cm]{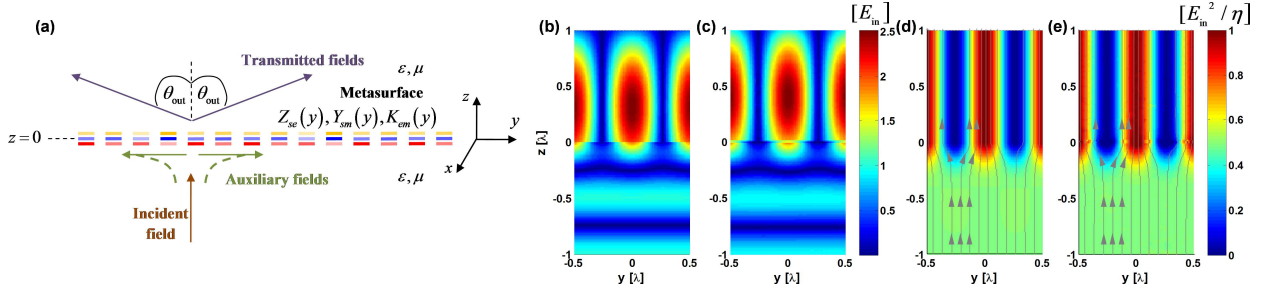}%
 \caption{\label{fig:beamSplitter}O-BMS for reflectionless wide-angle beam splitting. (a) Physical configuration. (b) Analytically predicted [Eqs. \eqref{equ:fields_above_beam_splitter} and \eqref{equ:fields_below_beam_splitter}] and (c) simulated electric field distribution $\left|\Re\left\{E_x\left(y,z\right)\right\}\right|$. (d) Analytically predicted and (e) simulated $z$-directed real power $P_z\left(y,z\right)=\frac{1}{2}\Re\left\{E_xH_y^*\right\}$; gray streamlines indicate flow of real power from bottom to top.}
 \end{figure*}

\begin{figure*}
 \includegraphics[width=16.5cm]{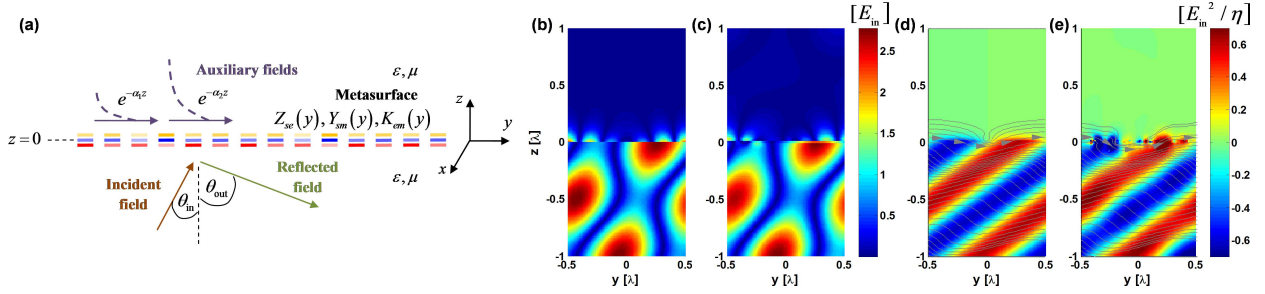}%
 \caption{\label{fig:perfectReflector}O-BMS for perfect reflection. (a) Physical configuration. (b) Analytically predicted [Eqs. \eqref{equ:fields_below_perfect_reflector} and \eqref{equ:fields_above_perfect_reflector}] and (c) simulated electric field distribution $\left|\Re\left\{E_x\left(y,z\right)\right\}\right|$. (d) Analytically predicted and (e) simulated $z$-directed real power $P_z\left(y,z\right)=\frac{1}{2}\Re\left\{E_xH_y^*\right\}$; gray streamlines indicate flow of real power from left to right.}
 \end{figure*}

For simplicity, we consider a two-dimensional configuration ($\partial/\partial x=0$) with transverse electric (TE) fields ($E_z=E_y=H_x=0$), where an O-BMS occupies the plane $z=0$, surrounded by homogeneous medium with permittivity $\epsilon$ and permeability $\mu$ [Figs. \ref{fig:beamSplitter}(a), \ref{fig:perfectReflector}(a)]. The O-BMS is characterized by its electric surface impedance $Z_{se}\left(y\right)$, magnetic surface admittance $Y_{sm}\left(y\right)$, and magnetielectric coupling coefficient $K_{em}\left(y\right)$. 
The wave impedance and wave number are given, respectively, by $\eta=\sqrt{\mu/\epsilon}$ and $k=\omega\sqrt{\mu\epsilon}$, assuming a time-harmonic dependency of $e^{j\omega t}$. 
The fields below ($z<0$) and above ($z>0$) the metasurface are denoted as $\left\{E_x^{<}\left(y,z\right), H_y^{<}\left(y,z\right), H_z^{<}\left(y,z\right)\right\}$ and $\left\{E_x^{>}\left(y,z\right), H_y^{>}\left(y,z\right), H_z^{>}\left(y,z\right)\right\}$, respectively.

The relation between the tangential fields on the bottom (${z\rightarrow0^-}$) and top (${z\rightarrow0^+}$) facets of the metasurface is given by the omega-type bianisotropic sheet transition conditions (O-BSTCs) \cite{Radi2013,Epstein2016_3}
\begin{equation}
	\begin{array}{l}
	\vspace{2pt}
	\frac{1}{2}\!\left(E_{x}^{+}\!+\!E_{x}^{-}\right)\!=\!-{{Z}_{se}}\!\left( H_{y}^{+}\!-\!H_{y}^{-} \right)\!-\!{{K}_{em}}\!\left( E_{x}^{+}\!-\!E_{x}^{-} \right)  \\
   \frac{1}{2}\!\left(H_{y}^{+}\!+\!H_{y}^{-}\right)\!=\!-\!{{Y}_{sm}}\!\left( E_{x}^{+}\!-\!E_{x}^{-} \right)\!+\!{{K}_{em}}\!\left( H_{y}^{+}\!-\!H_{y}^{-} \right)\!,
	\end{array}
\label{equ:BSTCs}
\end{equation}
where we define ${E_{x}^{-}\left(y\right)\!=\!\left.E_x^{<}\left(y,z\right)\right|_{z\rightarrow0^-}}$,  ${E_{x}^{+}\left(y\right)\!=\!\left.E_x^{>}\left(y,z\right)\right|_{z\rightarrow0^+}}$ (and analogously for $H_x^{\pm}$), and the $y$ dependency is omitted for brevity. 

As discussed in \cite{Epstein2016_3}, if the fields just below and above the metasurface \emph{locally} conserve the real power $P_z\left(y\right)$, namely,
\begin{equation}
\textstyle
P_z^{-}\!\left(y\right)\!=\!\frac{1}{2}\Re\!\!\left\{E_x^-H_y^{-*}\right\}
	\!=\!\frac{1}{2}\Re\!\!\left\{E_x^+H_y^{+*}\right\}\!=\!P_z^{+}\!\left(y\right)\!,
\label{equ:local_power_conservation}
\end{equation}
for each $y$ on the metasurface, then there exists a passive and lossless solution ($\Re\{Z_{se}\}=\Re\{Y_{sm}\}=\Im\{K_{em}\}=0$ \cite{Radi2013}) to the O-BSTCs, given by
\begin{equation}
\left\lbrace\!\!\!
\begin{array}{l}
	\vspace{3pt}
	{{K}_{em}}=\frac{1}{2}\frac{\Re \left\{ E_{x}^{+}H{{_{y}^{-}}^{*}}-E_{x}^{-}H{{_{y}^{+}}^{*}} \right\}}{\Re \left\{ \left( E_{x}^{+}-E_{x}^{-} \right){{\left( H_{y}^{+}-H_{y}^{-} \right)}^{*}} \right\}} \\ 
	\vspace{3pt}
 {{Y}_{sm}}=-j\left( \frac{1}{2}\Im \left\{ \frac{H_{y}^{+}+H_{y}^{-}}{E_{x}^{+}-E_{x}^{-}} \right\}-{{K}_{em}}\Im \left\{ \frac{H_{y}^{+}-H_{y}^{-}}{E_{x}^{+}-E_{x}^{-}} \right\} \right) \\ 
 \vspace{3pt}
 {{Z}_{se}}=-j\left( \frac{1}{2}\Im \left\{ \frac{E_{x}^{+}+E_{x}^{-}}{H_{y}^{+}-H_{y}^{-}} \right\}+{{K}_{em}}\Im \left\{ \frac{E_{x}^{+}-E_{x}^{-}}{H_{y}^{+}-H_{y}^{-}} \right\} \right). 
\end{array}
\right.
\label{equ:passive_lossless_design}
\end{equation}
In other words, the passive lossless O-BMS given by Eq. \eqref{equ:passive_lossless_design} will support the transformation between $\left\{E_x^<\left(y,z\right),H_y^<\left(y,z\right),H_z^<\left(y,z\right)\right\}$ and $\left\{E_x^>\left(y,z\right),H_y^>\left(y,z\right),H_z^>\left(y,z\right)\right\}$, provided these fields satisfy Eq. \eqref{equ:local_power_conservation}.

Our first goal is to design a passive and lossless O-BMS that splits a normally-incident plane wave to two equal-power plane waves propagating towards $\theta=\pm \theta_\mathrm{out}$ [Fig. \ref{fig:beamSplitter}(a)]. According to the derivation in Eqs. \eqref{equ:BSTCs}-\eqref{equ:passive_lossless_design}, to achieve this goal we need to stipulate the fields below and above the metasurface such that (i) they obey Maxwell's equations in the respective regions, (ii) they satisfy local power conservation [Eq. \eqref{equ:local_power_conservation}], and (iii) they coincide with the given incident and desirable transmitted fields in the far-field region.

To understand the challenge underlying these requirements, we examine the power profiles on the two facets of the metasurface. The desirable fields above the metasurface ($z>0$) are a superposition of two plane waves with equal amplitudes $E_\mathrm{out}$, namely,
\begin{equation}
\begin{array}{l}
	\vspace{3pt}
	\!\!\!\!\!\!E_x^>\!\!\left(y,z\right)\!\! =\!\! E_\mathrm{out}e^{-jkz\cos\theta_\mathrm{out}}\!\left(e^{-jky\sin\theta_\mathrm{out}}\!+\!e^{jky\sin\theta_\mathrm{out}}\right)\! \\
	\!\!\!\!\!\!H_y^>\!\!\left(y,z\right)\!\!=\!\!\frac{E_\mathrm{out}}{Z_\mathrm{out}}e^{-jkz\cos\theta_\mathrm{out}}\!\left(e^{-jky\sin\theta_\mathrm{out}}\!+\!e^{jky\sin\theta_\mathrm{out}}\right)\!\!, 
\end{array}\!
\label{equ:fields_above_beam_splitter}
\end{equation}
where $Z_\mathrm{out}=\eta/\cos\theta_\mathrm{out}$ is the TE wave impedance. The real power profile just above the metasurface ($z\rightarrow0^+$) is thus given by [Eq. \eqref{equ:local_power_conservation}]
\begin{equation}
	P_z^+\left(y\right) = 4\frac{\left|E_\mathrm{out}\right|^2}{Z_\mathrm{out}}\cos^2\left(ky\sin\theta_\mathrm{out}\right),
\label{equ:power_above_beam_splitter}
\end{equation}
spatially varying following a squared cosine. On the other hand, a simple plane wave normally-incident upon the metasurface from below ($z<0$) would form a uniform power profile on the bottom facet ($z\rightarrow0^-$).

Therefore, to match the power profiles on the two facets of the metasurface, auxiliary fields are required. The idea is to introduce suitable evanescent modes to the region below the metasurface such that 
their interference with the incident field would redistribute the (otherwise uniform) real power on the bottom facet $P_z^-\left(y\right)$, leading to the desirable squared-cosine modulation matching $P_z^+\left(y\right)$. Being evanescent, these auxiliary fields would not affect the performance in the far field, as required. 

One possible choice is to add surface waves propagating at opposite directions (towards $y\rightarrow\pm\infty$) on the bottom facet of the metasurface. 
With this choice, the fields below the metasurface ($z<0$) would be
\begin{equation}
\begin{array}{l}
	E_x^<\left(y,z\right) = E_\mathrm{in}e^{-jkz} 
	 +E_\mathrm{sw}e^{\alpha_\mathrm{sw}z}
	\left(e^{-2jky\sin\theta_\mathrm{out}}+e^{2jky\sin\theta_\mathrm{out}}\right) \\
	H_y^<\left(y,z\right) = \frac{E_\mathrm{in}}{Z_\mathrm{in}}e^{-jkz} 
	+j\frac{\alpha_\mathrm{sw}E_\mathrm{sw}}{k\eta}
	e^{\alpha_\mathrm{sw}z}\left(e^{-2jky\sin\theta_\mathrm{out}}+e^{2jky\sin\theta_\mathrm{out}}\right), 
\end{array}
\label{equ:fields_below_beam_splitter}
\end{equation}
where $E_\mathrm{in}$ is the (given) amplitude of the incident field, $Z_\mathrm{in}=\eta/\cos\theta_\mathrm{in}$ is the wave impedance ($\theta_\mathrm{in}=0$), $E_\mathrm{sw}$ is the amplitude of the surface waves, and $\alpha_\mathrm{sw}=k\sqrt{4\sin^2\theta_\mathrm{out}-1}\in\mathbb{R}$ is the surface wave decay constant. The latter definition implies that this design scheme is valid only for wide-angle beam splitting ($\theta_\mathrm{out}>30^\circ$), for which the auxiliary fields are evanescent.

Assuming the field amplitudes are related via $E_\mathrm{sw}=aE_\mathrm{in}$, $a\in\mathbb{R}$, the real power on the bottom facet of the metasurface ($z\rightarrow0^-$) will be given by
\begin{equation}
	P_z^-\left(y\right) = \frac{\left|E_\mathrm{in}\right|^2}{Z_\mathrm{in}}\left[1+2a\cos\left(2ky\sin\theta_\mathrm{out}\right)\right].
\label{equ:power_below_beam_splitter}
\end{equation}
Comparing Eq. \eqref{equ:power_above_beam_splitter} with Eq. \eqref{equ:power_below_beam_splitter} indicates that local power conservation Eq. \eqref{equ:local_power_conservation} may be achieved if we choose $E_\mathrm{out}=\frac{1}{\sqrt{2}}E_\mathrm{in}e^{-j\xi_\mathrm{out}}\sqrt{Z_\mathrm{out}/Z_\mathrm{in}}$ and $a=\frac{1}{2}$, where $\xi_\mathrm{out}$ is a possible uniform phase shift.

This step finalizes the design procedure. Fixing the amplitudes as prescribed allows evaluation of the fields everywhere in space via Eq. \eqref{equ:fields_below_beam_splitter} and Eq. \eqref{equ:fields_above_beam_splitter}, and in particular on the metasurface facets ($z\rightarrow0^\pm$). As these fields satisfy Maxwell's equations 
and local power conservation [Eq. \eqref{equ:local_power_conservation}], we may substitute them into Eq. \eqref{equ:passive_lossless_design} to assess the (passive lossless) O-BMS specifications that implement the required field transformation.

To verify our theory, we follow the outlined procedure and design an O-BMS that couples a normally-incident plane wave to two plane waves propagating towards $\theta=\pm\theta_\mathrm{out}=\pm 71.81^\circ$ with equal power, incurring a phase shift of $\xi_\mathrm{out}=145^\circ$. The design is implemented in ANSYS HFSS following the scheme introduced in \cite{Wong2016, Epstein2016_3}, based on three reactive sheets defined on two $5\mathrm{mil}$-thick Rogers RO3010 substrates 
\cite{SupMatOBMS}. In the design frequency $f=20\mathrm{GHz}$, the unit cell size is $\Delta y=\lambda/9.5$ and the metasurface thickness is $\lambda/59$. 

One period of the metasurface $\Lambda=\lambda/\sin\theta_\mathrm{out}=10\Delta y$ was simulated with periodic boundary conditions, indicating that $49.5\%$ of the incident power is coupled to each of the transmitted plane waves, while as little as $1\%$ is specularly reflected. The analytically-predicted field distributions $\left|\Re\{E_x\left(y,z\right)\}\right|$ [Fig. \ref{fig:beamSplitter}(b)] agree very well with the 
simulated ones [Fig. \ref{fig:beamSplitter}(c)], verifying that 
the normally-incident wavefront is transformed to two beams propagating towards $\pm71.81^\circ$, yielding a typical interference pattern above the metasurface. The field distribution clearly shows the auxiliary surface waves formed just below the metasurface, as stipulated in Eq. \eqref{equ:fields_below_beam_splitter}. 

The role of the auxiliary fields is further highlighted in Fig. \ref{fig:beamSplitter}(d) and (e), where the real part of the $z$-directed Poynting vector $P_z\left(y,z\right)$ is depicted. The normally-incident plane wave carries uniform power in the far field, but as the interference with the auxiliary surface wave becomes more significant near the O-BMS, the power is redistributed to match the power profile $P_z^+\left(y\right)$ of the splitted beam above the aperture. This is emphasized by the gray lines following the power flow, indicating the diversion of power towards the peaks (red) and away of the valleys (blue) of the sqaured cosine of Eq. \eqref{equ:power_above_beam_splitter}. We stress that the fields stipulated in Eqs. \eqref{equ:fields_above_beam_splitter}-\eqref{equ:power_above_beam_splitter} form an \emph{exact} solution to Maxwell's equations with the relevant boundary conditions, such that the passive lossless O-BMS design is strictly accurate.

The second functionality we consider is "perfect" engineered reflection \cite{Asadchy2016}. In this case, we desire to fully couple a plane wave with an angle of incidence $\theta_\mathrm{in}$ to a plane wave reflected towards $\theta_\mathrm{out}$ [Fig. \ref{fig:perfectReflector}(a)]. The fields below the metasurface can be thus written as
\begin{equation}
\begin{array}{l}
	E_x^<\left(y,z\right) = E_\mathrm{in}e^{-jkz\cos\theta_\mathrm{in}}e^{-jky\sin\theta_\mathrm{in}} 
	 -E_\mathrm{out}e^{jkz\cos\theta_\mathrm{out}}e^{-jky\sin\theta_\mathrm{out}}
	\\
	\vspace{2pt}
	H_y^<\left(y,z\right) = \frac{E_\mathrm{in}}{Z_\mathrm{in}}e^{-jkz\cos\theta_\mathrm{in}}e^{-jky\sin\theta_\mathrm{in}} 
	+\frac{E_\mathrm{out}}{Z_\mathrm{out}}e^{jkz\cos\theta_\mathrm{out}}e^{-jky\sin\theta_\mathrm{out}}, 
\end{array}
\label{equ:fields_below_perfect_reflector}
\end{equation}
where $E_\mathrm{in}$ and $E_\mathrm{out}$ are the amplitudes of the incident and reflected plane waves, respectively. 
To guarantee \emph{global} power conservation 
(integrated over the aperture), we set $E_\mathrm{out}=E_\mathrm{in}e^{-j\xi_\mathrm{out}}\sqrt{Z_\mathrm{out}/Z_\mathrm{in}}$, where $\xi_\mathrm{out}$ is a possible phase shift. 
Hence, the real power [Eq. \eqref{equ:local_power_conservation}] on the bottom facet of the metasurface ($z\rightarrow0^-$) is given by
\begin{equation}
	P_z^-\!\!\left(y\right)\!=\!\frac{\left|E_\mathrm{in}\right|^2}{2Z_\mathrm{in}}\!\!\left(\!\!\sqrt{\frac{Z_\mathrm{in}}{Z_\mathrm{out}}}\!-\!\sqrt{\frac{Z_\mathrm{out}}{Z_\mathrm{in}}}\right)\!\cos\left(ky\Delta_\mathrm{sin}\!+\!\xi_\mathrm{out}\right)\!,
\label{equ:power_below_perfect_reflector}
\end{equation}
where we define $\Delta_\mathrm{sin}=\sin\theta_\mathrm{out}-\sin\theta_\mathrm{in}$.

This power profile is even more intricate than the one 
examined before [Eq. \eqref{equ:power_above_beam_splitter}], as it changes signs along the metasurface plane: power is crossing the metasurface from bottom to top at one half of a period, whereas in the other half, power is crossing the metasurface from top to bottom. This means that to support such a field transformation, which deflects all the power incident from $\theta_\mathrm{in}$ towards $\theta_\mathrm{out}$, with a passive lossless O-BMS, it is required to conceive a mechanism which draws the power from below at one half-period and pushes it back to the bottom half-space at the other half-period.

Naively trying to stipulate zero fields on the top facet would lead, in general, to lossy and active design requirements \cite{Asadchy2016}, as it would violate local power conservation [Eq. \eqref{equ:local_power_conservation}]. However, 
introducing auxiliary evanescent modes above the metasurface can assist matching the desirable power profile of Eq. \eqref{equ:power_below_perfect_reflector} without any effect on the device functionality. The idea is to achieve the oscillating power profile via interference between two surface waves travelling on the top facet of the metasurface. As surface waves do not generally carry real power perpendicular to their propagation direction, it is required to introduce a $\pi/2$ phase shift between them. Correspondingly, we define the fields above the O-BMS as
\begin{equation}
\begin{array}{l}
	E_x^>\!\!\left(y,z\right)\!=\!E_\mathrm{sw,1}e^{-\alpha_1 z}e^{-jk_{t,1}y}\!+\!jE_\mathrm{sw,2}e^{-\alpha_2 z}e^{-jk_{t,2}y}
	\\
	H_y^>\!\!\left(y,z\right)\!=\!-j\frac{\alpha_1 E_\mathrm{sw,1}}{k\eta}e^{-\alpha_1 z}e^{-jk_{t,1}y} \!+\!\frac{\alpha_2 E_\mathrm{sw,2}}{k\eta}e^{-\alpha_2 z}e^{-jk_{t,2}y}, 
\end{array}
\label{equ:fields_above_perfect_reflector}
\end{equation}
where $E_\mathrm{sw,n}$, $\left|k_{t,n}\right|>k$, and $\alpha_n = \sqrt{k_{t,n}^2-k^2}>0$, are the amplitude, transverse wavenumber, and decay coefficient of the $n$th surface wave, respectively. 
Without loss of generality, we define $E_\mathrm{sw,2}=be^{-j\xi}E_\mathrm{sw,1}$ ($b\in\mathbb{R}$), yielding the following power profile on the top facet
\begin{equation}
	P_z^+\left(y\right) = \frac{b\left|E_\mathrm{sw,1}\right|^2}{2\eta}\frac{\alpha_2-\alpha_1}{k}\cos\left[\left(k_{t,2}-k_{t,1}\right)y+\xi\right].
\label{equ:power_above_perfect_reflector}
\end{equation}
Comparing Eq. \eqref{equ:power_above_perfect_reflector} with Eq. \eqref{equ:power_below_perfect_reflector} reveals that local power conservation [Eq. \eqref{equ:local_power_conservation}] may be achieved by setting $k_{t,2}=k_{t,1}+k\Delta_\mathrm{sin}$, $\xi=\xi_\mathrm{out}$, and ${b=\frac{k\eta \left|E_\mathrm{in}\right|^2}{\left(\alpha_2-\alpha_1\right)Z_\mathrm{in}\left|E_\mathrm{sw,1}\right|^2}\left(\sqrt{\frac{Z_\mathrm{in}}{Z_\mathrm{out}}}-\sqrt{\frac{Z_\mathrm{out}}{Z_\mathrm{in}}}\right)}$. 
The parameters $k_{t,1}$ and $E_\mathrm{sw,1}$ can be chosen at will, as long as $\left|k_{t,n}\right|>k$.

As for the beam-splitter application, once we achieve local power conservation the design procedure is completed. The stipulated fields in Eqs. \eqref{equ:fields_below_perfect_reflector} and \eqref{equ:fields_above_perfect_reflector} can be evaluated on the metasurface facets ($z\rightarrow 0^{\pm}$) and substituted into Eq. \eqref{equ:passive_lossless_design} to yield the O-BMS specifications.

To verify this scheme, we follow the procedure outlined in Eqs. \eqref{equ:fields_below_perfect_reflector}-\eqref{equ:power_above_perfect_reflector} to design a perfect reflector which fully couples a plane wave incident from $\theta_\mathrm{in}=0$ to a plane wave reflected towards $\theta_\mathrm{out}=71.81^\circ$, imposing a phase shift of $\xi_\mathrm{out}=90^\circ$; for simplicity, we choose $k_{t,1}=2k$, $E_\mathrm{sw,1}=E_\mathrm{in}$. One period of the O-BMS $\Lambda=\lambda/\Delta_\mathrm{sin}=10\Delta y$ is once more implemented in ANSYS HFSS using cascaded reactive sheets \cite{Wong2016,Epstein2016_3,SupMatOBMS}.

The analytically-predicted field distribution $\left|\Re\{E_x\left(y,z\right)\}\right|$ [Fig. \ref{fig:perfectReflector}(b)] agrees well with the simulated one [Fig. \ref{fig:perfectReflector}(c)], indicating that $98.5\%$ of the power is successfully coupled from the incident mode to the reflected one. The interference patterns of the incident and reflected fields below and the surface waves above the O-BMS are clearly observed. The $z$-directed Poynting vector $P_z\left(y,z\right)$ shown in Fig. \ref{fig:perfectReflector}(d) and (e) indicates that the auxiliary fields guarantee its continuity on the aperture, 
shifting power upwards (red) and downwards (blue) alternately. The gray power-flow streamlines demonstrate that, indeed, the interference between the auxiliary surface waves above the O-BMS forms the mechanism by which power can circulate between the bottom and top facets, establishing local power conservation. 

To conclude, we have presented a paradigm for synthesis of O-BMSs for prescribed beam manipulation, without requiring active or lossy components, and without any compromise on device performance. Our approach makes use of auxiliary fields to facilitate the local power conservation required for achieving reactive designs; being evanescent, these fields do not interfere with the device functionality in the far field. We have demonstrated this concept by designing O-BMSs for reflectionless beam splitting and perfect reflection, functionalities that were considered impossible to implement accurately with passive lossless metasurfaces. 

It should be stressed that the fields analytically stipulated using this approach are \emph{exact} solutions to Maxwell's equations with the 
O-BSTCs, and no approximation is used in the process. Importantly, no external sources are required to excite the auxiliary fields: they are self-consistently generated by the currents induced on the metasurface due to the incident fields \cite{SupMatOBMS}. Integrating this unorthodox approach into metasurface synthesis would allow accurate reactive solutions for many other challenging applications.

\end{document}